\documentclass[10pt,a4paper,onecolumn]{article}
\usepackage{marginnote}
\usepackage{graphicx}
\usepackage{xcolor}
\usepackage{authblk,etoolbox}
\usepackage{titlesec}
\usepackage{calc}
\usepackage{tikz}
\usepackage{hyperref}
\hypersetup{colorlinks,breaklinks=true,
            urlcolor=[rgb]{0.0, 0.5, 1.0},
            linkcolor=[rgb]{0.0, 0.5, 1.0}}
\usepackage{caption}
\usepackage{tcolorbox}
\usepackage{amssymb,amsmath}
\usepackage{ifxetex,ifluatex}
\usepackage{seqsplit}
\usepackage{xstring}

\usepackage{float}
\let\origfigure\figure
\let\endorigfigure\endfigure
\renewenvironment{figure}[1][2] {
    \expandafter\origfigure\expandafter[H]
} {
    \endorigfigure
}

\usepackage{fixltx2e} 
\usepackage[
  backend=biber,
]{biblatex}
\bibliography{paper.bib}

\let\textttOrig=\texttt
\def\texttt#1{\expandafter\textttOrig{\seqsplit{#1}}}
\renewcommand{\seqinsert}{\ifmmode
  \allowbreak
  \else\penalty6000\hspace{0pt plus 0.02em}\fi}

\makeatletter
\let\href@Orig=\href
\def\href@Urllike#1#2{\href@Orig{#1}{\begingroup
    \def\Url@String{#2}\Url@FormatString
    \endgroup}}
\def\href@Notdoi#1#2{\def\tempa{#1}\def\tempb{#2}%
  \ifx\tempa\tempb\relax\href@Urllike{#1}{#2}\else
  \href@Orig{#1}{#2}\fi}
\def\href#1#2{%
  \IfBeginWith{#1}{https://doi.org}%
  {\href@Urllike{#1}{#2}}{\href@Notdoi{#1}{#2}}}
\makeatother

\newlength{\cslhangindent}
\newlength{\csllabelwidth}
\newenvironment{CSLReferences}[3] 
 {
  \setlength{\parindent}{0pt}
  \ifodd #1 \everypar{\setlength{\hangindent}{\cslhangindent}}\ignorespaces\fi
  \ifnum #2 > 0
  \setlength{\parskip}{#2\baselineskip}
  \fi
 }%
 {}
\usepackage{calc}

\usepackage[top=3.5cm, bottom=3cm, right=1.5cm, left=1.0cm,
            headheight=2.2cm, reversemp, includemp, marginparwidth=4.5cm]{geometry}

\titleformat{\section}
  {\normalfont\sffamily\Large\bfseries}
  {}{0pt}{}
\titleformat{\subsection}
  {\normalfont\sffamily\large\bfseries}
  {}{0pt}{}
\titleformat{\subsubsection}
  {\normalfont\sffamily\bfseries}
  {}{0pt}{}
\titleformat*{\paragraph}
  {\sffamily\normalsize}

\usepackage{fancyhdr}
\makeatletter
\let\ps@plain\ps@fancy
\fancyheadoffset[L]{4.5cm}
\fancyfootoffset[L]{4.5cm}

\definecolor{linky}{rgb}{0.0, 0.5, 1.0}

\newtcolorbox{repobox}
   {colback=red, colframe=red!75!black,
     boxrule=0.5pt, arc=2pt, left=6pt, right=6pt, top=3pt, bottom=3pt}

\patchcmd{\@maketitle}{center}{flushleft}{}{}
\patchcmd{\@maketitle}{center}{flushleft}{}{}
\patchcmd{\@maketitle}{\LARGE}{\LARGE\sffamily}{}{}
\def\maketitle{{%
  
  \AB@maketitle}}
\makeatletter
\renewcommand\AB@affilsepx{ \protect\Affilfont}
\renewcommand\AB@affilnote[1]{{\bfseries #1}\hspace{3pt}}
\renewcommand{\affil}[2][]%
   {\newaffiltrue\let\AB@blk@and\AB@pand
      \if\relax#1\relax\def\AB@note{\AB@thenote}\else\def\AB@note{#1}%
        \setcounter{Maxaffil}{0}\fi
        \begingroup
        \let\href=\href@Orig
        \let\texttt=\textttOrig
        \let\protect\@unexpandable@protect
        \def\thanks{\protect\thanks}\def\footnote{\protect\footnote}%
        \@temptokena=\expandafter{\AB@authors}%
        {\def\\{\protect\\\protect\Affilfont}\xdef\AB@temp{#2}}%
         \xdef\AB@authors{\the\@temptokena\AB@las\AB@au@str
         \protect\\[\affilsep]\protect\Affilfont\AB@temp}%
         \gdef\AB@las{}\gdef\AB@au@str{}%
        {\def\\{, \ignorespaces}\xdef\AB@temp{#2}}%
        \@temptokena=\expandafter{\AB@affillist}%
        \xdef\AB@affillist{\the\@temptokena \AB@affilsep
          \AB@affilnote{\AB@note}\protect\Affilfont\AB@temp}%
      \endgroup
       \let\AB@affilsep\AB@affilsepx
}
\makeatother

\renewcommand\Affilfont{\sffamily\small\mdseries}
\ifnum 0\ifxetex 1\fi\ifluatex 1\fi=0 
  \usepackage[T1]{fontenc}
  \usepackage[utf8]{inputenc}

\else 
  \ifxetex
    \usepackage{mathspec}
    \usepackage{fontspec}

  \else
    \usepackage{fontspec}
  \fi
  \defaultfontfeatures{Ligatures=TeX,Scale=MatchLowercase}

\fi
\IfFileExists{upquote.sty}{\usepackage{upquote}}{}
\IfFileExists{microtype.sty}{%
\usepackage{microtype}
\UseMicrotypeSet[protrusion]{basicmath} 
}{}

\usepackage{hyperref}
\hypersetup{unicode=true,
            pdftitle={Learning source-aware representations of music in a discrete latent space},
            pdfborder={0 0 0},
            breaklinks=true}
\usepackage{longtable,booktabs}

\let\addcontentslineOrig=\addcontentsline
\def\addcontentsline#1#2#3{\bgroup
  \let\texttt=\textttOrig\addcontentslineOrig{#1}{#2}{#3}\egroup}
\let\markbothOrig\markboth
\def\markboth#1#2{\bgroup
  \let\texttt=\textttOrig\markbothOrig{#1}{#2}\egroup}
\let\markrightOrig\markright
\def\markright#1{\bgroup
  \let\texttt=\textttOrig\markrightOrig{#1}\egroup}

\usepackage{graphicx,grffile}
\makeatletter
\def\maxwidth{\ifdim\Gin@nat@width>\linewidth\linewidth\else\Gin@nat@width\fi}
\def\maxheight{\ifdim\Gin@nat@height>\textheight\textheight\else\Gin@nat@height\fi}
\makeatother
\setkeys{Gin}{width=\maxwidth,height=\maxheight,keepaspectratio}
\IfFileExists{parskip.sty}{%
\usepackage{parskip}
}{
\setlength{\parindent}{0pt}
\setlength{\parskip}{6pt plus 2pt minus 1pt}
}
\ifx\paragraph\undefined\else
\let\oldparagraph\paragraph
\renewcommand{\paragraph}[1]{\oldparagraph{#1}\mbox{}}
\fi
\ifx\subparagraph\undefined\else
\let\oldsubparagraph\subparagraph
\renewcommand{\subparagraph}[1]{\oldsubparagraph{#1}\mbox{}}
\fi

\title{Learning source-aware representations of music in a discrete
latent space}

        \author[1]{Jinsung Kim\footnote{co-first author}}
          \author[1]{Yeong-Seok Jeong\footnote{co-first author}}
          \author[2]{Woosung Choi}
          \author[3]{Jaehwa Chung}
          \author[1]{Soonyoung Jung\footnote{corresponding author}}
    
      \affil[1]{Korea University}
      \affil[2]{Queen Mary University of London}
      \affil[3]{Korea National Open University}
  \date{\vspace{-7ex}}

\begin{document}
\maketitle

\marginpar{

  \begin{flushleft}
  \sffamily\small

  \vspace{2mm}

  \par\noindent\hrulefill\par

  \vspace{2mm}

  \vspace{2mm}
  {\bfseries License}\\
  Authors of papers retain copyright and release the work under a Creative Commons Attribution 4.0 International License (\href{http://creativecommons.org/licenses/by/4.0/}{\color{linky}{CC BY 4.0}}).

  \vspace{4mm}
  {\bfseries In partnership with}\\
  \vspace{2mm}
  \includegraphics[width=4cm]{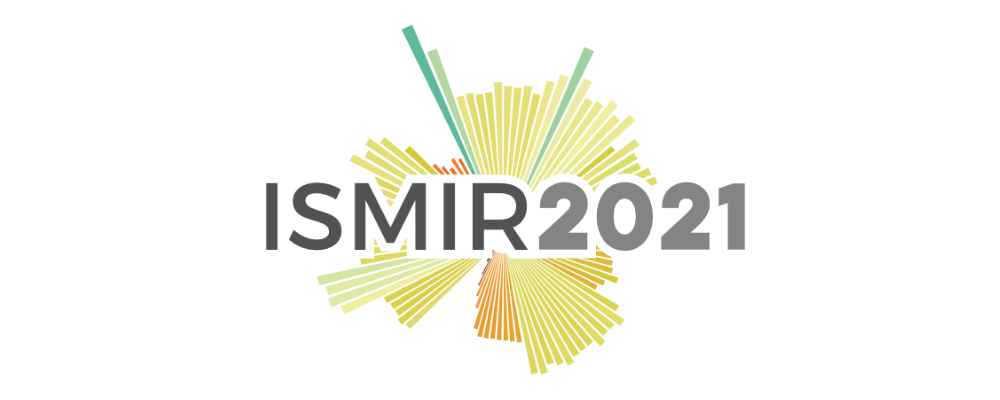}

  \end{flushleft}
}

\hypertarget{abstract}{%
\section{Abstract}\label{abstract}}

In recent years, neural network based methods have been proposed as a 
method that can generate representations from music, but they are not human
readable and hardly analyzable or editable by a human. To address this
issue, we propose a novel method to learn source-aware latent
representations of music through Vector-Quantized Variational
Auto-Encoder(VQ-VAE). We train our VQ-VAE to encode an input mixture
into a tensor of integers in a discrete latent space, and design them to
have a decomposed structure which allows humans to manipulate the latent
vector in a source-aware manner. This paper also shows that we can
generate bass lines by estimating latent vectors in a discrete space.

\hypertarget{introduction}{%
\section{Introduction}\label{introduction}}

Humans have a unique auditory cortex, allowing proficiency at hearing
sounds, and sensitiveness in frequencies. For this reason, humans can
differentiate individual sources in an audio signal by capturing
acoustic characteristics unique to each source (e.g., timbre and
tessitura). Moreover, experts or highly skilled composers are able to
produce sheet music for different instruments, just by listening to the
mixed audio signal. Meanwhile, trained orchestras or bands can reproduce
the original music by playing the transcribed scores. However, if an
unskilled transcriber lacks the ability to distinguish different music
sources, no matter how good the performers are, they cannot recreate the
original music. This procedure resembles the encoder-decoder concept,
widely used in the machine learning field; the transcriber is an encoder
(Hung et al., 2020; Kim \& Bello, 2019), and the orchestra/band is a
decoder(Klejsa et al., 2019; Ren et al., 2020). Motivated by this
analogy, this paper proposes a method that aims to learn source-aware
decomposed audio representations for a given music signal. Instead of
predicting high-level representations such as scores or MIDI, our method
aims to predict discrete representations proposed in (Oord et al.,
2017). To the best of our knowledge, numerous methods have been proposed
for audio representation, yet no existing works have learned decomposed
music representations in a discrete latent space.

\hypertarget{related-work}{%
\section{Related work}\label{related-work}}

For automatic speech recognition, (Baevski et al., 2020; Sadhu et al.,
2021) used Transformer (Vaswani et al., 2017)-based models to predict
quantized latent vectors. From this approach, they trained their models
to understand linguistic information in human utterance. (Ericsson et
al., 2020; Mun et al., 2020) learn voice style representations from
human voice information. They applied learned representations for speech
separation and its enhancement. Several studies have applied contrastive
learning, used in representation learning@{[}niizumi2021byol; Wang \&
Oord (2021){]} for computer vision.

However, the goal of this paper is different from the above audio
representation researches. We aim to learn decomposed representations
through instruments' categories. In this work, we train a model through
source separation to learn decomposed representations. In section
``Experiment'' , we show that we can easily manipulate latent
representations for various applications such as source separation and
music synthesis. Source Separation tasks have been studied both in music
source separation and on speech enhancement tasks. Within the generating
perspective, they can be categorized into two groups. The first group
attempts to generate masks that is multiplied with the input audio to
acquire the target source (Chien et al., 2017; Jansson et al., 2017).
The second group aims to directly estimate a raw audio or spectrogram
(Choi et al., 2020; Kameoka et al., 2019; Samuel et al., 2020).

This method can generate audio samples directly when we have a prior
distribution of representation. Many studies have proposed methods based
on Variational Auto-Encoder (VAE)(Kameoka et al., 2019) or U-Net (Choi
et al., 2020; Samuel et al., 2020; Yuan et al., 2019) for source
separation. The U-Net-based models usually show high performance in the
source separation task. However, some studies have pointed out the
fundamental limitation of U-Nets; the skip connections used in U-Nets
may lead to weakening the expressive power of encoded representations
(Yuan et al., 2019). Therefore, we choose a VAE-based model to extract
meaningful representation from the input audio.

\hypertarget{proposed-methods}{%
\section{Proposed Methods}\label{proposed-methods}}

When a mixture audio is given, we aim to learn quantized representations
that can be decomposed into \(n\) vectors, where \(n\) denotes the
number of sources. We denote the given source set as
\(S=\{s_i\}_{i=1}^n\), where \(s_i \in S\) is the \(i^{th}\) source. We
formulate the mixture audio \(\mathcal{M}(S)\) as follows:
\[ \mathcal{M}(S) = \sum_{i}^{n} s_i\]

In the following section, we describe our method for learning decomposed
latent representations. Figure 1 describes the overall idea of the
proposed model.

\begin{figure}
\centering
\includegraphics[width=1\textwidth,height=\textheight]{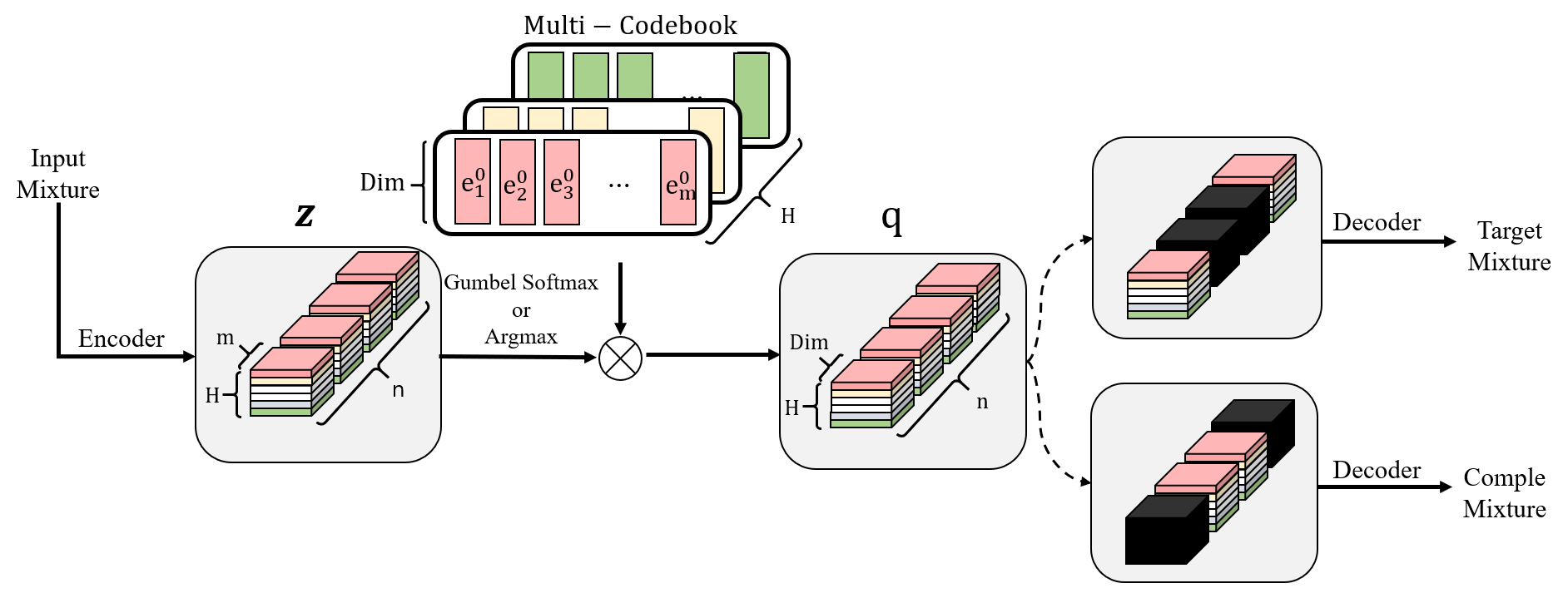}
\caption{Illustration of the proposed method.}
\end{figure}

\hypertarget{latent-quantization}{%
\subsection{Latent Quantization}\label{latent-quantization}}

We quantize latent vectors by adopting the vector quantization method
proposed in (Oord et al., 2017). In this original VQ-VAE (Oord et al.,
2017), straight-through estimation was used, but instead we adopted
Gumbel softmax estimation (Jang et al., 2016) where the softmax is done
over all encoder outputs z.
\[ y_{i,j} = \frac{\exp((z_{i,j} + g_j) / \tau ) }{\sum_{k}^{m}\exp((z_{i,k} + g_k)/\tau) }\]
while \(\tau\) is a non-negative temperature, \(g\) are categorical
distribution from \(n= -log(-log(u))\), and \(u\) are uniform samples
from uniform distribution \(\mathcal{U}(0,1)\).

During training, Gumbel softmax estimation stochastically maps the
representation of the encoder output to the discrete latents. In other
words, \(q_i\) is a weighted sum of all discrete latents. This way it is
easier to compute the gradients and can simply implement the
reparameterization trick that approximates the categorical
distributions. However, during inference we deterministically choose one
discrete latent in the codebook.

\[ q_i =\begin{cases}
   [e_1, e_2, ... e_m] \otimes y_i &\quad \text{if train step} \\
   e^*                   &\quad \text{if inference step}
\end{cases}\] where \(q_i\) is the quantized representation of the
\(i\)-th source, \(e\) are the discrete latents in the codebook, \(z\)
is the output of encoder, and \(e^*=e_{\operatorname{argmax}(y_i)}\).

\hypertarget{multi-latent-quantization}{%
\subsection{Multi-latent Quantization}\label{multi-latent-quantization}}

One limitation of the latent quantization approach is the restricted
expressive power compared to continuous latent space approaches. To
increase the number of source representations, we could have simply
increased the number of elements in a codebook. Instead, we use a more
memory-efficient way to increase expressive power. We use multiple
codebooks and extract each quantized vector \(q_i^{(h)}\) from the
\(h^{th}\) codebook. By concatenating these quantized vectors 
$q_i^{(h)}$, we construct quantized vector \(q_i\).

\[ q_i=[q_i^{(1)}, ..., q_i^{(H)}], (h \in [1,H])\]

Through this approach, the number of available source representations
increases exponentially with the total number of codebooks.

\hypertarget{task-definition}{%
\subsection{Task definition}\label{task-definition}}

When the input mixture \(\mathcal{M}(S)\) is given, we want to obtain
decomposed and quantized representations \({q_i}^{n}_{i=1}\). We assume
that each decomposed representation vector \(q_{i}\) can fully represent
each \(s_i\).

If we select some representations from \([q_1, ..., q_n]\), they also
have to represent \(s_i\) that are chosen. For example, if we select
\(q_1\), \(q_2\) and \(q_4\), these are the representation of
\(\mathcal{M}(\{s_1, s_2, s_4\})\). We apply this
\textit{seletive source reconstruction task} to our model, where we aim
to minimize the \textit{selective reconstruction loss}. The selective
source reconstruction loss can be formulated, as follows:

\[ \mathcal{L}_{select} = {\| \mathcal{M}(S')  - \hat{\mathcal{M}}' \|}_1^1\]
where \(\hat{\mathcal{M}}'\) is estimated audio through decoder network
from selected representation.

When calculating the gradients through the selective latent
reconstruction loss, there is no gradient for unselected
representations. This lack of gradients is inefficient to train the
model. To prevent this problem, we train our model with the
\textit{complement source reconstruction task}, where we aim to minimize
the \textit{complement loss}. The complement loss
\(\mathcal{L}_{compl}\) is defined as follows:
\[ \mathcal{L}_{compl} = {\| \mathcal{M}(S'')  - \hat{\mathcal{M}}''\|}_1^1\]
where \(\hat{\mathcal{M}}''\) is estimated audio through decoder network
from unselected representation.

We also conduct the STFT loss as auxiliary loss defined as follows:
\[\mathcal{L}_{STFT} =
\| \operatorname{STFT}(\mathcal{M}) - \operatorname{STFT}(\hat{\mathcal{M}}) \|_1^1\]
We apply it to both \(\mathcal{L}_{select}\) and
\(\mathcal{L}_{compl}\).

\hypertarget{experiment}{%
\section{Experiment}\label{experiment}}

\hypertarget{dataset}{%
\subsection{Dataset}\label{dataset}}

We use the MUSDB18 dataset (Rafii et al., 2017). It contains a total of
150 tracks. We divided the training set into 86 tracks for training, 14
tracks for validation. In MUSDB18, each track has a sampling rate of
44100 and each mixture has four stereo sources(vocal, bass drum, and
other instruments).

\hypertarget{training-setting}{%
\subsection{Training Setting}\label{training-setting}}

Given an input audio, we trained models to predict the target audio,
following the selective source condition. The input mixture audio is a
summation of randomly selected sources from different tracks, and the
target sources are randomly selected from the input mixture's sources.
When computing the STFT loss, We set the number of FFTs to 2048, and hop
length to 512. We trained models using the Adam optimizer (Kingma \& Ba,
2014) and set the learning rate to 0.0001.

\hypertarget{results}{%
\subsection{Results}\label{results}}

To validate our method, we visualize the result of decomposed
representations using t-SNE (Van der Maaten \& Hinton, 2008). After
training with the MUSDB18 training dataset, we obtained decomposed
representations of single-source tracks in the MUSDB18 test dataset.
Then we apply t-SNE to the set of representations as shown in Figure 2.
In Figure 2, each color means different sources and the dots are the
decomposed representations. It can be examined that the latent vectors
from the same sources tend to be clustered even though there is no
constraint about the classification. It indicates that our method has
learned source-aware representations.

\begin{figure}
\centering
\includegraphics[width=0.6\textwidth,height=\textheight]{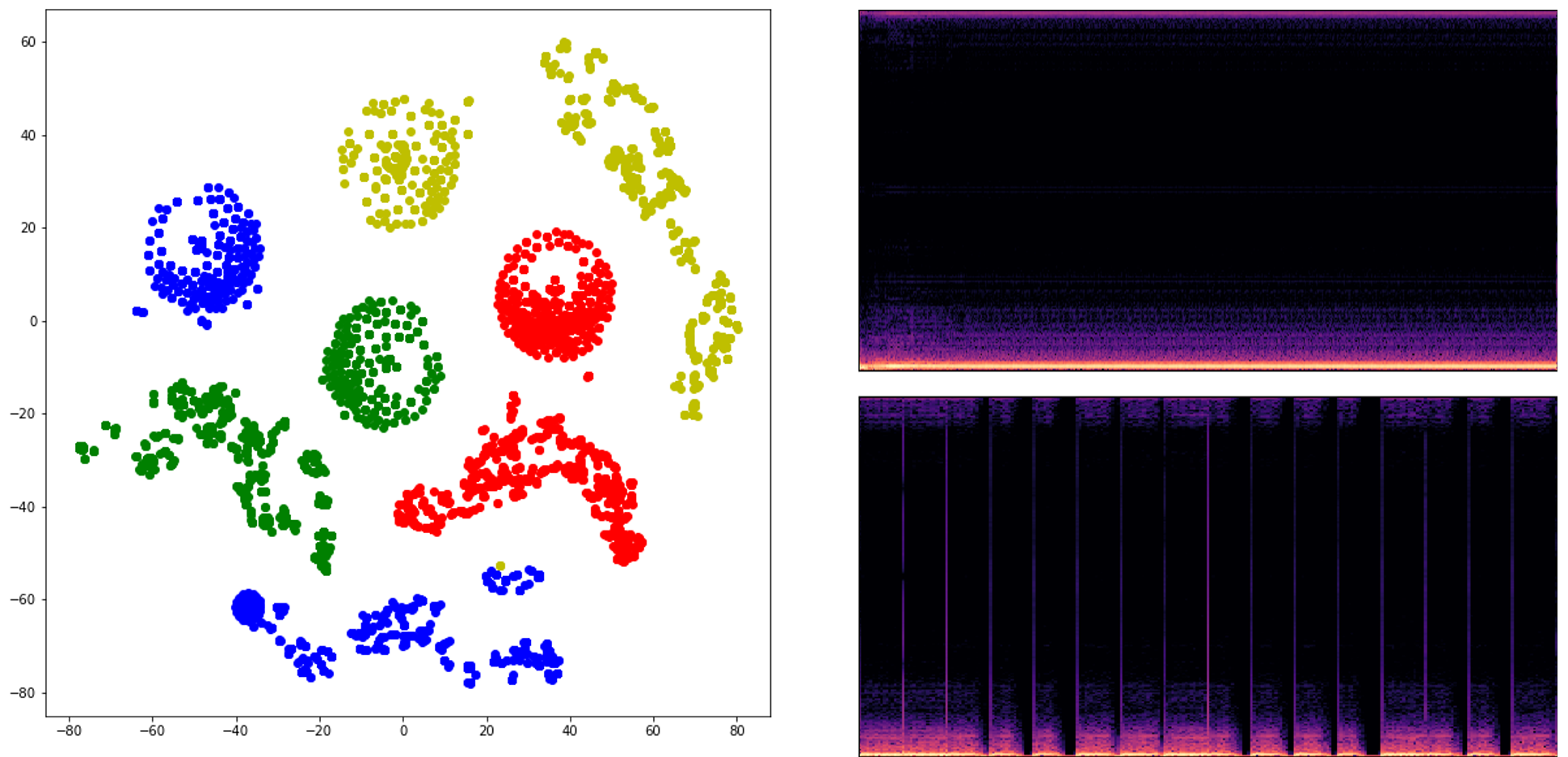}
\caption{tSNE visualization of quantized vectors in multi-codebook(left)
and bass generation result(right)}
\end{figure}

We also conduct an experiment using methods without the STFT and
complement loss, introduced in Section ``Proposed Methods'' to compare
the effects of them. To this end, we first separate sources from
mixtures in the MUSDB18 test dataset using each model. Then, we measure
Source-to-Distortion Ratio (SDR) @{[}vincent2006performance{]},
following MUSDB18 benchmark to evaluate each models.

\begin{longtable}[]{@{}cccccl@{}}
\toprule
& vocals & bass & drums & other & Avg\tabularnewline
\midrule
\endhead
proposed & 1.270 & 1.761 & 1.403 & 0.812 & 1.311\tabularnewline
w/o STFT & 1.546 & 1.026 & 1.480 & 1.069 & 1.280\tabularnewline
w/o comple & 0.996 & -0.031 & 1.458 & 0.576 & 0.749\tabularnewline
Demucs (Défossez et al., 2019) & 6.84 & 7.01 & 6.86 & 4.42 &
6.28\tabularnewline
\bottomrule
\end{longtable}

Compared to Demucs proposed in (Défossez et al., 2019), the SDR scores
of our model are low. We argue that low SDR scores are caused by the
absence of U-skip connections (i.e., skip-connections between encoder
and decoder used in U-Net (Ronneberger et al., 2015)). Many
state-of-the-art models Choi et al. (2020) have adopted U-skip
connections. However, ours does not employ them because ours focused on
the feasibility of training of decomposed and quantized latent vectors
rather than the source separation itself. Due to the absence of
skip-connection, the decoder might struggle to generate high-quality
audio. Also, the vector quantization might generate blurrier contents as
discussed in (Oord et al., 2017), which might degrade the SDR scores.
However, even though our method has generated a low SDR score, it has a
considerable potential to be applied to other tasks, as discussed in the
Conclusion section.

\hypertarget{conclusion}{%
\section{Conclusion}\label{conclusion}}

This paper explores learning decomposed representations for musical
signals. We propose novel training methods (i.e., selective source
reconstruction and complementary source reconstruction) and a
VQ-VAE-based model. To validate our approaches, we visualize the latent
representation through the t-SNE algorithm and perform two experiments.
The visualized representation shows that the latent representations of
sources are decomposed into different spaces. The bass generation task
shows that the decoder can generate bass lines via new prior. We
consider that our model can be used in other music processing tasks. For
example, our model, which represents discrete representations of input
audio, can be adopted in music compression tasks. In addition, the
characteristics of the model generating decomposed audio representation
for each source is appropriate for the music sheet transcription task.
We plan to design a decoder that can generate high-quality audio. This
can be applied to real-world audio in future work. This methodology can
generate a decomposed representation of the various sounds of the real
world. As a result, it can be implemented to various tasks such as audio
generation and audio event detection, and localization.

\hypertarget{acknowledgements}{%
\section{Acknowledgements}\label{acknowledgements}}

This research was supported by Basic Science Research Program through
the National Research Foundation of Korea(NRF) funded by the Ministry of
Education(NRF-2021R1A6A3A03046770). This work was also supported by the
National Research Foundation of Korea(NRF) grant funded by the Korea
government(MSIT)(No.~NRF-2020R1A2C1012624, NRF-2021R1A2C2011452).

\hypertarget{references}{%
\section*{References}\label{references}}
\addcontentsline{toc}{section}{References}

\hypertarget{refs}{}
\begin{CSLReferences}{1}{0}
\leavevmode\hypertarget{ref-baevski2020wav2vec}{}%
Baevski, A., Zhou, H., Mohamed, A., \& Auli, M. (2020). wav2vec 2.0: A
framework for self-supervised learning of speech representations.
\emph{arXiv Preprint arXiv:2006.11477}.

\leavevmode\hypertarget{ref-chien2017variational}{}%
Chien, J.-T., Kuo, K.-T., \& others. (2017). Variational recurrent
neural networks for speech separation. \emph{18th ANNUAL CONFERENCE OF
THE INTERNATIONAL SPEECH COMMUNICATION ASSOCIATION (INTERSPEECH 2017),
VOLS 1-6: SITUATED INTERACTION}, 1193--1197.

\leavevmode\hypertarget{ref-choi2020lasaft}{}%
Choi, W., Kim, M., Chung, J., \& Jung, S. (2020). LaSAFT: Latent source
attentive frequency transformation for conditioned source separation.
\emph{arXiv Preprint arXiv:2010.11631}.

\leavevmode\hypertarget{ref-defossez:2019}{}%
Défossez, A., Usunier, N., Bottou, L., \& Bach, F. (2019). Music source
separation in the waveform domain. \emph{arXiv Preprint
arXiv:1911.13254}.

\leavevmode\hypertarget{ref-ericsson2020adversarial}{}%
Ericsson, D., Östberg, A., Zec, E. L., Martinsson, J., \& Mogren, O.
(2020). Adversarial representation learning for private speech
generation. \emph{arXiv Preprint arXiv:2006.09114}.

\leavevmode\hypertarget{ref-hung2020transcription}{}%
Hung, Y.-N., Wichern, G., \& Roux, J. L. (2020). \emph{Transcription is
all you need: Learning to separate musical mixtures with score as
supervision}. \url{http://arxiv.org/abs/2010.11904}

\leavevmode\hypertarget{ref-jang2016categorical}{}%
Jang, E., Gu, S., \& Poole, B. (2016). Categorical reparameterization
with gumbel-softmax. \emph{arXiv Preprint arXiv:1611.01144}.

\leavevmode\hypertarget{ref-jansson2017orcid}{}%
Jansson, A., Humphrey, E., Montecchio, N., Bittner, R., Kumar, A., \&
Weyde, T. (2017). ORCID: 0000-0001-8028-9905 (2017). Singing voice
separation with deep u-net convolutional networks. \emph{18th
International Society for Music Information Retrieval Conference},
23--27.

\leavevmode\hypertarget{ref-kameoka2019supervised}{}%
Kameoka, H., Li, L., Inoue, S., \& Makino, S. (2019). Supervised
determined source separation with multichannel variational autoencoder.
\emph{Neural Computation}, \emph{31}(9), 1891--1914.

\leavevmode\hypertarget{ref-kim2019adversarial}{}%
Kim, J. W., \& Bello, J. P. (2019). Adversarial learning for improved
onsets and frames music transcription. \emph{arXiv Preprint
arXiv:1906.08512}.

\leavevmode\hypertarget{ref-kingma2014adam}{}%
Kingma, D. P., \& Ba, J. (2014). Adam: A method for stochastic
optimization. \emph{arXiv Preprint arXiv:1412.6980}.

\leavevmode\hypertarget{ref-klejsa2019high}{}%
Klejsa, J., Hedelin, P., Zhou, C., Fejgin, R., \& Villemoes, L. (2019).
High-quality speech coding with sample RNN. \emph{ICASSP 2019-2019 IEEE
International Conference on Acoustics, Speech and Signal Processing
(ICASSP)}, 7155--7159.

\leavevmode\hypertarget{ref-mun2020sound}{}%
Mun, S., Choe, S., Huh, J., \& Chung, J. S. (2020). The sound of my
voice: Speaker representation loss for target voice separation.
\emph{ICASSP 2020-2020 IEEE International Conference on Acoustics,
Speech and Signal Processing (ICASSP)}, 7289--7293.

\leavevmode\hypertarget{ref-oord2017neural}{}%
Oord, A. van den, Vinyals, O., \& Kavukcuoglu, K. (2017). Neural
discrete representation learning. \emph{Proceedings of the 31st
International Conference on Neural Information Processing Systems},
6309--6318.

\leavevmode\hypertarget{ref-rafii2017musdb18}{}%
Rafii, Z., Liutkus, A., Stöter, F.-R., Mimilakis, S. I., \& Bittner, R.
(2017). \emph{MUSDB18-a corpus for music separation}.

\leavevmode\hypertarget{ref-ren2020popmag}{}%
Ren, Y., He, J., Tan, X., Qin, T., Zhao, Z., \& Liu, T.-Y. (2020).
Popmag: Pop music accompaniment generation. \emph{Proceedings of the
28th ACM International Conference on Multimedia}, 1198--1206.

\leavevmode\hypertarget{ref-unet}{}%
Ronneberger, O., Fischer, P., \& Brox, T. (2015). U-net: Convolutional
networks for biomedical image segmentation. \emph{International
Conference on Medical Image Computing and Computer-Assisted
Intervention}, 234--241.

\leavevmode\hypertarget{ref-sadhu2021wav2vec}{}%
Sadhu, S., He, D., Huang, C.-W., Mallidi, S. H., Wu, M., Rastrow, A.,
Stolcke, A., Droppo, J., \& Maas, R. (2021). Wav2vec-c: A
self-supervised model for speech representation learning. \emph{arXiv
Preprint arXiv:2103.08393}.

\leavevmode\hypertarget{ref-9053513}{}%
Samuel, D., Ganeshan, A., \& Naradowsky, J. (2020). Meta-learning
extractors for music source separation. \emph{ICASSP 2020 - 2020 IEEE
International Conference on Acoustics, Speech and Signal Processing
(ICASSP)}, 816--820.
\url{https://doi.org/10.1109/ICASSP40776.2020.9053513}

\leavevmode\hypertarget{ref-van2008visualizing}{}%
Van der Maaten, L., \& Hinton, G. (2008). Visualizing data using t-SNE.
\emph{Journal of Machine Learning Research}, \emph{9}(11).

\leavevmode\hypertarget{ref-vaswani2017attention}{}%
Vaswani, A., Shazeer, N., Parmar, N., Uszkoreit, J., Jones, L., Gomez,
A. N., Kaiser, Ł., \& Polosukhin, I. (2017). Attention is all you need.
\emph{Advances in Neural Information Processing Systems}, 5998--6008.

\leavevmode\hypertarget{ref-wang2021multi}{}%
Wang, L., \& Oord, A. van den. (2021). Multi-format contrastive learning
of audio representations. \emph{arXiv Preprint arXiv:2103.06508}.

\leavevmode\hypertarget{ref-yuan2019skip}{}%
Yuan, W., Wang, S., Li, X., Unoki, M., \& Wang, W. (2019). A skip
attention mechanism for monaural singing voice separation. \emph{IEEE
Signal Processing Letters}, \emph{26}(10), 1481--1485.

\end{CSLReferences}

\end{document}